\documentstyle[aps,prl,multicol,floats,epsf]{revtex}

\begin{document}
\draft
\wideabs{

\title{
Electrical Resistivity Anisotropy from Self-Organized One-Dimensionality 
in High-Temperature Superconductors}

\author{Yoichi Ando, Kouji Segawa, Seiki Komiya, and A. N. Lavrov}
\address{Central Research Institute of Electric Power 
Industry, Komae, Tokyo 201-8511, Japan}

\date{\today}
\maketitle

\begin{abstract}
We investigate the manifestation of 
stripes in the in-plane resistivity anisotropy in untwinned 
single crystals of La$_{2-x}$Sr$_x$CuO$_4$ ($x$ = 0.02--0.04) 
and YBa$_2$Cu$_3$O$_y$ ($y$ = 6.35--7.0).  
It is found that both systems show strongly temperature-dependent 
in-plane anisotropy in the lightly hole-doped region and that 
the anisotropy in YBa$_2$Cu$_3$O$_y$ {\it grows} with decreasing $y$ 
below $\sim$6.60 despite the decreasing orthorhombicity, which gives 
most direct evidence that electrons self-organize into a 
macroscopically anisotropic state.  
The transport is found to be easier along the direction of the 
spin stripes already reported, 
demonstrating that the stripes are intrinsically conducting in cuprates.
\end{abstract}

}

\narrowtext

The mechanism of the high-temperature superconductivity is not settled 
15 years after its discovery, mostly because it is unclear how best to 
describe the strongly-correlated electrons in the high-$T_c$ cuprates.  
It has recently been discussed 
\cite{Tranquada,Yamada,Mook1,Matsuda,Hunt,Mook2,MR,Noda,Zhou,%
EKZ,Kivelson,Carlson,Zaanen,CastroNeto} that the electrons 
in cuprates self-organize into quasi-one-dimensional stripes, 
which might bring a paradigm shift in our way of understanding 
the two-dimensional (2D) electronic system in the cuprates.  
Though the self-organization is interesting in itself, 
the conventional wisdom suggests that the stripes are destructive 
to the superconductivity, because charge ordering would normally 
lead to an insulating state.  
However, there are intriguing theoretical proposals 
\cite{EKZ,Kivelson,Carlson} that stripes can instead be responsible 
for the {\it occurrence} of the high-temperature superconductivity 
if they are conducting and meandering --- properties that have never 
been clearly demonstrated before.  

In this Letter, we report novel in-plane transport anisotropy 
in the cuprates, which gives direct evidence for the conducting 
charge stripes in these materials; 
the temperature dependence and the magnitude of the anisotropy 
strongly suggest the stripes to be meandering and forming an 
electronic liquid crystal \cite{Kivelson}.  
The evidence is shown for two representative materials of the 
high-temperature superconductors, 
La$_{2-x}$Sr$_x$CuO$_4$ (LSCO) and YBa$_2$Cu$_3$O$_y$ (YBCO), 
highlighting the universality of the charge-stripe phenomenon 
in the cuprates.  
Most notably, the data for YBCO indicate that the charge stripes 
govern the transport in samples with $T_c$ of as high as 50 K, 
demonstrating that theories of high-temperature superconductivity 
should inevitably consider the self-organization of the electrons 
as an integral part. 

It is fair to say that the majority of the researchers today 
believe that the stripes are irrelevant to the superconductivity.  
This general belief comes not only from the conventional wisdom 
mentioned above but also from the current experimental situation, 
which can be summarized as follows: 
(i) strong evidence for the charge stripe order has been 
reported \cite{Tranquada} only for Nd-doped LSCO, 
where the superconductivity is strongly suppressed; 
(ii) in other superconducting cuprates, evidence 
\cite{Yamada,Mook1,Matsuda,Hunt,Mook2,MR} is reasonably strong for 
{\it spin} stripes, but not at all conclusive for {\it charge} stripes.  
Therefore, to really establish the charge stripe as a new paradigm, 
it is necessary to clarify whether {\it charges} truly self-organize 
into stripes in superconducting cuprates other than Nd-doped LSCO.  
Moreover, since the stripe order in cuprates may just be a more 
or less standard charge density wave or spin density wave, 
it is important to clarify whether the charge stripes in cuprates, 
if exist, are intrinsically conducting; 
in fact, none of the previous works 
\cite{Tranquada,Yamada,Mook1,Matsuda,Hunt,Mook2,MR,Noda,Zhou} 
directly show the relation 
between the stripe order and metallic conduction. 

It is instructive to note that there is now little doubt about the 
existence of conducting charge stripes in the 2D electron gas in 
high Landau levels \cite{Fradkin}, for which a clear transport anisotropy 
\cite{Lilly1,Lilly2} has been considered to be convincing evidence.  
If one could observe a similar resistivity anisotropy in the 
nearly square CuO$_2$ planes in the cuprates, that would most 
convincingly establish the existence of the conducting charge stripes.  
Recent neutron scattering experiments \cite{Matsuda} have found static 
spin stripes that are unidirectional and extend along the $a$-axis of the 
orthorhombic crystalline lattice in LSCO with $x \le 0.05$ at low 
temperatures, which motivates us to search for the resistivity 
anisotropy in the lightly-doped LSCO.  
It has also been suggested \cite{Mook1} that the stripes are unidirectional 
and run along the $b$-axis in YBCO, which is the other system 
studied here; however, existence of the CuO chains \cite{Jorgensen}, 
which also run along the $b$-axis and are possibly conducting, may 
well contribute to the in-plane resistivity anisotropy in YBCO and thus a 
careful doping-dependence study is necessary to distinguish 
the stripe conductivity in the CuO$_2$ planes from the chain conductivity.  
Since these weakly-orthorhombic cuprates (orthorhombicity is only up 
to 1.7\%) \cite{Matsuda,Jorgensen} naturally develop twin structures 
in single crystals, it is necessary to detwin the crystals to make an 
in-plane anisotropy measurement. 

The LSCO crystals with $x$ = 0.02--0.04 used for this study are 
grown by the traveling-solvent floating-zone method \cite{mobility}.  
After determining the crystallographic axes using the X-ray Laue 
analysis, the LSCO crystals are cut into a platelet shape and are 
detwinned \cite{Lavrov} under a uniaxial pressure.  
Although it is difficult to observe the twin structure in the 
$ab$ face for LSCO, we found that the twins cause a clear 
washboard-like structure on the polished $ac$/$bc$ face, 
which is observable with both an optical microscope and a 
scanning electron microscope (Fig. 1) \cite{twin}.
Discovery of this feature enabled us to select only those 
LSCO crystals that are almost perfectly detwinned for the 
present measurement, while well-detwinned LSCO crystals were 
never available before.  
The YBCO crystals with $y$ = 6.35--7.0 are prepared and 
characterized as is described elsewhere \cite{60K}.  
With YBCO crystals, we can easily select perfectly detwinned 
crystals by looking at the $ab$ face with a polarized-light optical 
microscope.

\begin{figure}
\epsfxsize=8cm
\centerline{\epsffile{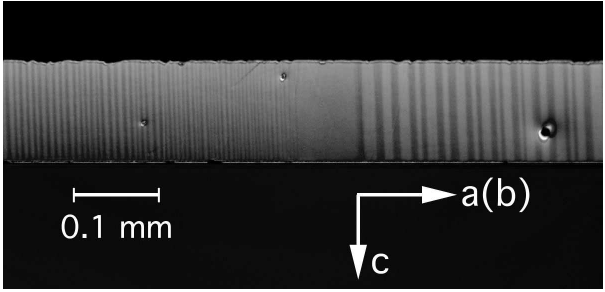}}
\vspace{0.2cm}
\caption{Twin pattern on the $ac$/$bc$ face of an as-grown LSCO crystal 
observed with an optical microscope. 
The contrast is caused by a washboard-like modulation on the surface
\protect\cite{twin}, which is confirmed by a scanning 
electron microscope.}
\label{fig1}
\end{figure}

The resistivity along the $a$-axis ($\rho_a$) and the $b$-axis ($\rho_b$) 
is measured by a straightforward four-terminal method with the accuracy 
and reproducibility of better than 10\% \cite{mobility,60K}. 
For LSCO, it turned out that 10\% accuracy was not good enough to determine 
the $ab$ anisotropy; to obtain most reliable data on the anisotropy 
of LSCO, we detwin the {\it same} piece of crystal {\it three-times} from 
different directions; namely, the crystals are initially detwinned to 
be $a$-axis oriented along the current direction to measure $\rho_a$, 
then to be $b$-axis oriented to measure $\rho_b$ without changing the 
contacts (thus the difference comes purely from the change 
in the axis orientation), and lastly to be $a$-axis oriented 
again to confirm that the first $\rho_a$ data are recovered. 
Figures 2(a)-(c) show the temperature dependences of $\rho_a$ and 
$\rho_b$ obtained this way for $x$ = 0.02, 0.03, and 0.04, and 
Fig. 2(d) shows the resulting in-plane anisotropy $\rho_b/\rho_a$ 
\cite{note1}.  
At high temperatures $\rho_b/\rho_a$ is nearly 1, which is consistent 
with the weak orthorhombicity (only 1.5\%) \cite{Matsuda}; 
however, what is unusual is that $\rho_b/\rho_a$ grows rapidly with 
decreasing temperature below $\sim$150 K.  
It is useful to note that at room temperature $\rho_b$ is slightly 
smaller than $\rho_a$ for all $x$ values and 
the anisotropy switches at lower temperature.  
Both the rapid growth and the switching of the anisotropy cannot be 
explained if one assumes that the anisotropy is due simply to the 
orthorhombicity of the crystal.  
If it is not due to the orthorhombicity, we must conclude that a 
self-organization of the electrons is responsible for the observed 
anisotropy behavior, because there is no other external source to cause 
the anisotropy in LSCO \cite{note2}.  
Since the preferred direction of the stripes is the $a$-axis 
\cite{Matsuda}, the observed in-plane anisotropy ($\rho_a < \rho_b$) 
below $\sim$100 K \cite{mag} indicates that the stripes are 
intrinsically conducting at finite temperature; 
this conclusion does not contradict the localization behavior at 
low temperatures, because charges in low dimensional systems 
easily localize in the presence of disorder.

\begin{figure}
\epsfxsize=1.0\columnwidth
\centerline{\epsffile{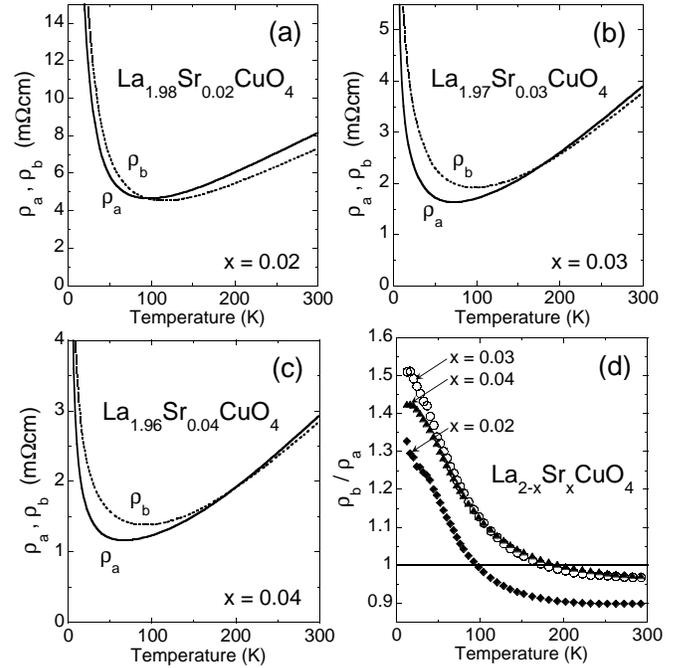}}
\vspace{0.2cm}
\caption{Anisotropic resistivity and the anisotropy ratio of 
lightly-doped LSCO. $T$ dependences of $\rho_a$ and $\rho_b$ 
are shown for $x$ = 0.02 (a), 0.03 (b), and 0.04 (c). 
Note that the spin stripes have been found \protect\cite{Matsuda} 
to run along the $a$-axis, along which the resistivity becomes smaller 
at low temperatures.}
\label{fig2}
\end{figure}

For YBCO, $\rho_a$ and $\rho_b$ are measured on different crystals, 
but they are respectively measured on at least 3 crystals for 
each composition and the data are confirmed to be very 
reproducible within 5\%, as is demonstrated elsewhere \cite{60K}.  
Representative data sets of $\rho_a(T)$ and $\rho_b(T)$ for four 
values of $y$, including the upper ($y$ = 7.0) and lower ($y$ = 6.35) 
extremes of this study, are shown in Figs. 3(a)-(d), 
and the anisotropy ratio, $\rho_a/\rho_b$, is shown in Fig. 4(a) 
for selected $y$.  
Previously, the in-plane resistivity anisotropy of YBCO has been 
reported \cite{Takenaka} only for $y > 6.6$, where $\rho_a/\rho_b$ 
monotonically decreases with decreasing $y$; 
this is consistent with the idea \cite{Takenaka} that the anisotropy 
near optimum doping is caused (at least partly) by an additional 
conductivity of the CuO chains which are progressively destroyed 
\cite{Jorgensen} with decreasing oxygen content.  However, when we 
extend the region of $y$ to lower values, the result is surprising; 
as can be seen in Fig. 4(a), $\rho_a/\rho_b$ at low temperatures turns 
out to grow with decreasing $y$ for $y < 6.6$, and the temperature 
dependence of $\rho_a/\rho_b$ becomes similar to that of 
$\rho_b/\rho_a$ for LSCO in the lightly-doped region.  
Since Fig. 4(a) is rather complicated, 
the evolution of $\rho_a/\rho_b$ in the $y$ vs. $T$ plane is 
transparently depicted in Fig. 4(b) with a color map.  
One can see in Fig. 4(b) that with decreasing $y$ down to $\sim$6.6 
the anisotropy gradually weakens, which is likely to reflect the 
diminishing contribution from the chains; however, the anisotropy 
starts to grow below $y \simeq$ 6.6 at low temperatures, causing a 
novel peak at the lower left corner where $\rho_a/\rho_b$ amounts 
to 2.5.

\begin{figure}
\epsfxsize=1.0\columnwidth
\centerline{\epsffile{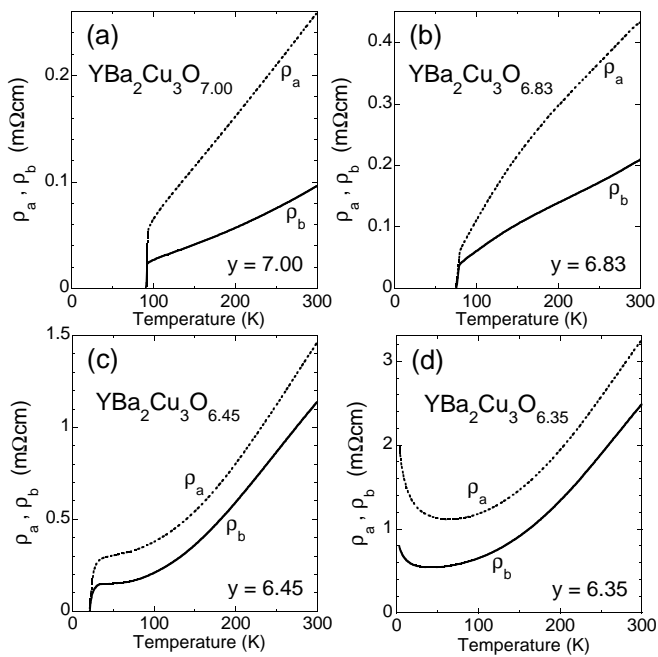}}
\vspace{0.2cm}
\caption{Representative data sets of $\rho_a(T)$ and $\rho_b(T)$ for 
YBCO at selected $y$.  
The $y$ values shown are 7.00 (a), 6.83 (b), 6.45 (c), and 6.35 (d).
In non-superconducting samples at $y$ = 6.35 (d), the anisotropy 
does not disappear even though the CuO chains are destroyed.}
\label{fig3}
\end{figure}

Because the orthorhombicity is gradually diminished \cite{Jorgensen} 
as the CuO chains are destroyed with decreasing $y$, 
the observed growth of $\rho_a/\rho_b$ for $y < 6.6$ cannot be simply 
due to the orthorhombicity nor the chain conductivity; 
therefore, we are forced to admit that the growth of $\rho_a/\rho_b$ with 
decreasing $y$ is caused by a self-organization of the two-dimensional 
electrons in the CuO$_2$ planes into an anisotropic system, 
which becomes stronger as the carrier concentration is reduced.  
Note that a clear in-plane resistivity anisotropy is observed even 
in a non-superconducting YBCO at $y$ = 6.35, where the orthorhombicity 
is about to disappear \cite{Jorgensen} (the measured lattice constants 
are $a$ = 3.871 \AA \ \ and $b$ = 3.861 \AA, 
giving only 0.26\% orthorhombicity).  
Most likely, the remaining short fragments of the chains 
(that are aligned to the $b$-axis upon detwinning) set a non-isotropic 
environment in which the electrons in the CuO$_2$ planes find the 
preferred orientation for their self-organization.  

We note that there is strong evidence that the CuO chains in YBCO are 
metallic at $y \simeq$ 7.0 \cite{Hussey,Bonn} and thus they certainly 
contribute to the anisotropy in fully oxygenated samples; however, 
the chain conductivity is expected to be quickly 
suppressed with decreasing $y$ because of the extreme sensitivity of 
one-dimensional (1D) systems to defects (even at $y$=6.90, each chain 
contains 10\% of defects and would normally be insulating). 
Therefore, it is possible that already near optimum doping the 
conductivity anisotropy is partly due to the (chain-induced) anisotropy 
in CuO$_2$ planes, which is actually suggested 
by a recent microwave measurement \cite{Bonn}. 
In heavily underdoped YBCO, the chain fragments are 
usually no longer than 10 unit cells even when an ordered ``ortho-II" 
phase is formed \cite{Haugerud}; thus, the ortho-II phase is not 
expected to cause any noticeable conductivity contribution through the 
chains.  In fact, the region of $y$ where the ortho-II phase is most 
stable (6.5 - 6.6) shows the {\it smallest} in-plane anisotropy.

It should be remarked that the anisotropy ratio only reaches 2.5 in 
the self-organized phase, 
which seems too small for a `quasi-1D' stripe phase; 
in the ordinary quasi-1D systems, such as Bechgaard salts, 
the conductivity along the 1D chains is usually a factor 
of 100 larger than that along the next best conducting direction.  
Thus, the simplistic picture of the rigid 1D stripes carrying 
current in the cuprates is clearly not valid.  On the other hand, the 
much smaller anisotropy and its strong temperature dependence 
observed here are actually consistent with the behavior \cite{Fradkin} 
of an electronic liquid crystal \cite{Kivelson}, 
where transverse fluctuations (quantum meandering) of the stripes are 
significant; 
it is useful to note that another self-organized stripe system, 
2D electrons in high Landau levels, also shows \cite{Lilly2} 
rather small anisotropy ratio of up to 7 with a strong 
temperature dependence \cite{note3}.  
It is intriguing that the electronic liquid crystals are predicted 
\cite{Kivelson} to be either a high-temperature superconductor or a 
two-dimensional anisotropic metal.  
The stripe phase of the lightly-doped cuprates 
seems to be consistent with the latter, and the YBCO data for 
$y < 6.6$ strongly suggest that the superconductivity with $T_c$ of 
at least 50 K is occurring in an electronic liquid crystal.  
If the charge transport at higher doping is also governed by 
the fluctuating stripes (as is suggested by the nearly 
doping-independent hole mobility \cite{mobility}), 
it may actually be the case that the whole phenomenon of 
high-temperature superconductivity is realized as a ground 
state of an electronic liquid crystal.  

In summary, we have found that the in-plane resistivity anisotropy 
in untwinned single crystals of LSCO and YBCO gives evidence for 
conducting charge stripes in these systems.  
The temperature dependence and the rather small magnitude 
of the anisotropy bear strong similarities to the nematic \cite{Fradkin} 
charge stripes in the 2D electron gas in high Landau Levels, suggesting 
that the electronic liquid crystals are likely to be realized in 
the cuprates.  Moreover, the signature of an electronic liquid crystal 
is observed in YBCO with $T_c$ of up to 50 K, which demonstrates that 
the electron self-organization is not a minor phenomenon in some 
extreme of the phase diagram but is rather an integral part of the 
physics in the cuprates.

We thank S. A. Kivelson for fruitful discussions and valuable comments.

%
\medskip
\vfil

\newpage
\widetext

\begin{figure}
\epsfxsize=15cm
\centerline{\epsffile{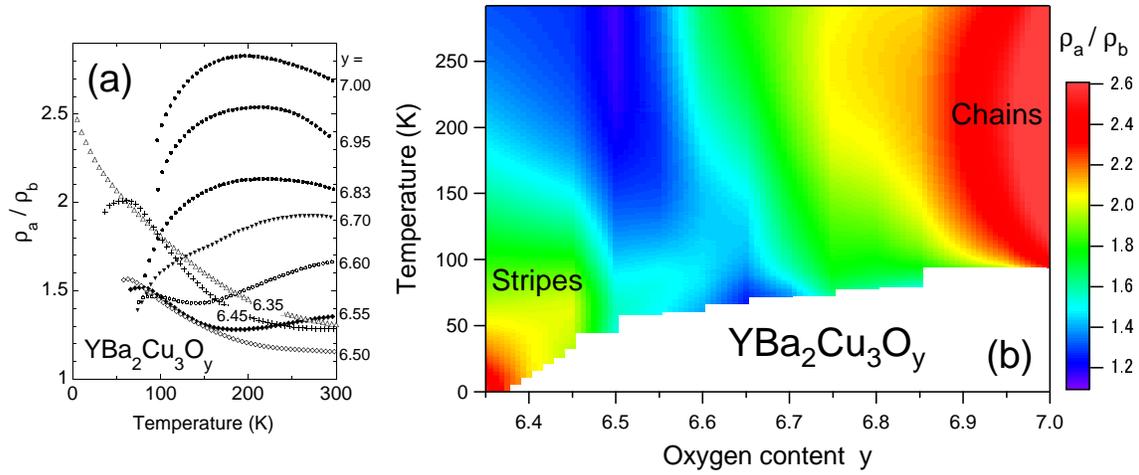}}
\vspace{0.2cm}
\caption{(a) Temperature dependences of $\rho_a/\rho_b$ for selected $y$.  
(b) Evolution of $\rho_a/\rho_b$ in the $y$ vs. $T$ plane. 
The white region corresponds to the superconducting state. 
The CuO chains cause a peak at $y$ = 7.0, which is gradually diminished 
as the chains are destroyed with decreasing $y$; 
on the other hand, a growth of $\rho_a/\rho_b$ with further decreasing $y$, 
observable for $y < 6.60$, signals the self-organization of the electrons 
into charge stripes. 
The anisotropy ratio at $y$ = 6.35, 6.45, 6.50, 6.55, 6.60, 6.65, 6.70, 
6.75, 6.80, 6.83, 6.95, and 7.00 are the actual data, and linear 
interpolations are employed to generate the color map.}
\label{fig4}
\end{figure}


\begin{references}
\bibitem{twin}
Across the twin boundary, the neighboring domains are orthorhombically 
distorted in opposite ways, and the polished surface appears to follow 
the direction of the crystallographic axis in each domain; this causes 
a washboard-like surface modulation.
\bibitem{note1}
Since we could not confirm the perfection of the second detwinning 
with the contacts on the sample, these $\rho_b/\rho_a$ data might 
slightly underestimate the true anisotropy; 
however, the $\rho_b$ data shown here agree with the data 
obtained separately on $\rho_b$-measurement-dedicated samples.
\bibitem{note2}
It is worth noting that in Figs. 2(a)-(c) the positions of the minimum 
in the resistivity curves, which marks the onset of charge localization 
\cite{mobility}, are different between the $a$ and $b$ directions for 
all $x$ values; such a behavior is not expected in an ordinary anisotropic 
metal, where the carriers in each band have an anisotropic effective 
mass with a single scattering rate.
\bibitem{mag}
Though the spin stripes are observed only below $\sim$70 K by neutrons 
\cite{Matsuda}, the magnetic susceptibility $\chi$ develops anisotropy 
from much higher temperature \cite{Lavrov}, as does the resistivity.  
However, the Dzyaloshinskii-Moriya interaction makes the anisotropy 
in $\chi(T)$ complicated, negating a direct comparison 
to the anisotropy in $\rho(T)$.
\bibitem{note3}
The anisotropy ratio for this system was initially reported to be 
order of 100 \cite{Lilly1}, which later turned out to be due to a 
geometric effect \cite{Lilly2}.
\end{references}
\end{document}